# Room temperature infrared photodetectors with hybrid structure based on 2D materials

Tiande Liu(刘天德)[1,†], Lei Tong(童磊)[1,†], Xinyu Huang(黄鑫宇)[1], Lei Ye(叶镭)[1,*]

[1] School of Optical and Electronic Information, Huazhong University of Science and Technology, Wuhan, Hubei 430074, China

[†] These author contributed equally

[*] Corresponding authors

Lei Ye: leiye@hust.edu.cn

**Abstract:** Two-dimensional (2D) materials, such as graphene, transition metal dichalcogenides (TMDs), black phosphorus (BP) and related derivatives, have attracted great attention due to their advantages of flexibility, strong light-matter interaction, broadband absorption and high carrier mobility, and have become a powerful contender for next-generation infrared photodetectors. However, since the thickness of two-dimensional materials is on the order of nanometers, the absorption of two-dimensional materials is very weak, which limits the detection performance of 2D materials-based infrared photodetector. In order to solve this problem, scientific researchers have tried to use optimized device structures to combine with two-dimensional materials for improving the performance of infrared photodetector. In this review, we review the progress of room temperature infrared photodetectors with hybrid structure based on 2D materials in recent years, focusing mainly on 2D-nD (n = 0, 1, 2) heterostructures, the integration between 2D materials and on-chip or plasmonic structure. Finally, we summarize the current challenges and point out the future development direction.

**Keywords:** two dimensional materials; heterostructure; infrared photodetectors;

**PACS:** 68.65.-k, 74.78.Fk, 85.60.Gz

## 1. Introduction

Infrared (IR) photodetectors can convert infrared light signals into easily detectable electrical signals, which are widely used in imaging, communications, medical and night vision applications at present.[1, 2] However, in order to obtain better detecting performance, the traditional infrared photodetectors based on the III-V or II-VI semiconductors have to work at low temperatures to reduce dark currents.[3] Because of the necessity of large refrigeration equipment to realize low operating temperature, it leads to the greatly increase of the manufacturing cost and limitation of the application, making it difficult to achieve miniaturization, flexibility and portability.

In order to circumvent this disadvantage, new type semiconductors are proposed to use in the application of IR photodetectors, which can achieve low current at room temperature and high photocurrent. In recent years, two-dimensional materials with an atomic layer thickness, are verified to overcome the shortcomings induced by traditional semiconductors used in traditional IR detection due to their excellent electrical, optical and mechanical properties.[4-13] For example, graphene is the most studied two-dimensional materials with semi-metal, high mobility, broad spectrum response, flexibility, and tunable Fermi level properties,[14, 15] suggesting that graphene can be fabricated into flexible infrared photodetectors with broad spectrum absorption, ultra-fast response.[16, 17] In addition, transition metal dichalcogenides materials $MX_2$ (M: transition metal atoms, X=S, Se, Te), which have the properties of strong light interactions and tunable bandgap varying with the number of layers, are also suitable as photosensitive materials for infrared photodetectors. Black phosphorus shows a direct bandgap property with the range of 0.3-1.3 eV, leading to its absorption spectrum covering the near-infrared, mid-infrared and far-infrared regions, even extending to the terahertz region.[18] Especially, wide spectral response range coupled with its unique anisotropic crystal structure, promises the potential of black phosphorus for infrared polarized photodetectors.[19] Although 2D materials have various excellent and unique properties for the applications of photodetector, they cannot avoid such a problem that such thin thickness cannot absorb more light. For example, the thickness of monolayer graphene is 0.35 nm and absorbs only 2.3% of light.[20] Although the bandgap of monolayer TMDs is a direct bandgap, the whole absorption is also less than 10%.[21]

In order to solve this problem, it is necessary to combine two-dimensional materials with other strong light absorbing materials, or to integrate with some structures that can enhance light absorption. For example, the band gap of quantum dots and nanoparticles with strong light absorption can be tuned by changing its size,[22, 23] which is very suitable as a photosensitive layer for infrared photodetectors. Carbon nanotubes have excellent broadband absorption characteristics,[24] which is also very suitable for forming heterostructures with two-dimensional materials. Integrating two-dimensional materials with waveguides or cavities can enhance the photon absorption of 2D materials to improve detection performance. In addition, this integration also reveals the compatibility of two-dimensional materials with existing optoelectronic technologies. The surface plasmon resonance effect not only enhances light absorption but also selectively absorbs light by designing different shapes and sizes,[25] therefore the spectral response of these designed photodetectors can be extended to infrared or terahertz region that cannot be absorbed by intrinsic two-dimensional material, due to the surface plasmon resonance effect.

In this review, we review the development of infrared photodetectors with mixed-dimensional structure between two-dimensional materials and other low-dimensional materials, focusing on quantum dots, single-wall carbon nanotubes, and 2D vertical or lateral heterostructures composed of different two-dimensional materials. Moreover, we also discuss the integration of two-dimensional materials and on-chip structures, including waveguides and cavities. We also introduce the application of surface plasmon resonance in infrared photodetectors based on two-dimensional materials. Finally, we summarize the faced challenges currently in infrared photodetectors with hybrid structures based on two-dimensional materials and point out the future directions for further development.

## 2. Heterostuctures

### 2.1 Quantum dots on 2D materials

In order to meet the demand for high-speed response of infrared photodetectors, the materials used to fabricate the detectors is often required to have high carrier mobility to reduce the transit time of photogenerated carriers in conducting channel. The carrier mobility of graphene is as high as 60,000 $cm^2V^{-1}s^{-1}$,[26] which can fully meet this demand. However, the optical response of photodetectors based on intrinsic graphene is only a few tens of mA/W,[27, 28] which is caused by weak light absorption due to its atomic-scale thickness. Compared with graphene, monolayer molybdenum disulfide ($MoS_2$) has a direct band gap, so the absorption is stronger than that of graphene, but still not enough. Moreover, the band gap of molybdenum disulfide is 1.8 eV, and the corresponding absorption wavelength range is in the visible light range, leading to the limitation of directly fabrication for an infrared photodetector.[21, 29] The problems of weak light absorption caused by the ultra-thin thickness and the unsatisfactory photoresponse range due to broad band gap, have seriously hindered the development of infrared photodetectors based on two-dimensional materials. To solve this problem, scientific researchers have tried many methods. It is a good attempt to combine two-dimensional materials with quantum dots.[16, 30-39] Because quantum dots have stronger light absorption relative to 2D materials, and their absorption wavelength range can be tuned by changing its size,[22, 40] and it can efficiently and selectively absorb photons and convert them into photogenerated electron-hole pairs. Therefore, the hybrid 2D-QDs structure is promising in the field of infrared detection technology.

Konstantatos et al. demonstrated a hybrid graphene-PbS quantum dots phototransistor with ultrahigh gain of $10^8$ and responsivity of $5\times10^7$ $AW^{-1}$.[32] The schematic diagram of a hybrid graphene-PbS quantum dots phototransistor is shown in Figure 1a. Monolayer or bilayer graphene was exfoliated with tape on a $Si/SiO_2$ substrate, and then an 80-nm PbS quantum dots film was prepared by low-cost spin casting method. PbS QDs thin film acts as light-sensitive material due to the characteristics of strong and selective light absorption, while graphene acts as a conduction channel owing to its extremely high carrier mobility. The energy band diagram at the interface of graphene and PbS QDs is shown in Figure 1b. On account of the work function mismatch between graphene and PbS QDs, the band bending occurs at the interface. PbS QDs thin film absorbs light to create photogenerated electron-hole pairs, which are then separated by an built-in electric field at the interface that induced by band bending. The Dirac point ($V_D$) will be drifted to higher backgate voltage due to illumination. For $V_{BG}<V_D$, holes are transferred from PbS QDs to graphene channel, and electrons are still trapped in the PbS QDs. The resistance of graphene decreases and carrier transport in the channel is hole-dominated. For $V_{BG}>V_D$, holes are transferred to graphene and recombined with electrons induced by the backgate, so the resistance of graphene increases and carrier transport in the channel is electron-dominated.

After the photogenerated holes are transferred to the graphene channel, they are driven by the source-drain voltage ($V_{SD}$) to reach the drain. The linear dependence of the photocurrent on $V_{SD}$ is shown in Figure 1c. In order to achieve charge conservation in graphene, holes are continuously replenished from the source. Due to the long charge trapping time of the PbS QDs and the high mobility of the graphene,[26, 41] holes in the channel can circulate multiple times during quantum dots trapped photogenerated electrons, resulting in ultrahigh photoconductive gain. The relationship between the responsivity and the applied back-gate voltage is shown in Figure 1d, which shows that the photoconductive gain of the photodetector can be adjusted by $V_{BG}$. For $V_{BG}<V_D$, the measured responsivity can reach up to $5\times10^7$ A/W at $V_{BG}$=-20 V; when $V_{BG}$=4 V, the responsivity reduces to zero. This back-gate-tunable gain is significant so that the required gain can be adjusted based on the photosignal intensity that needs to be detected.

Similarly, Zheng et al. demonstrated an ambipolar broadband photodetector based on N, S codecorated graphene-PbS quantum dot heterostructure.[42] N, S codecorated graphene greatly reduces the p-type doping of graphene in air. The gate voltage required to modulate the Dirac point and the carrier type is significantly reduced, thereby achieving gate tuneable mutual transformation of positive and negative photoresponse. The device exhibits excellent detection performance with responsivity of $10^4$ A/W and specific detectivity of $10^{12}$ Jones at 1550 nm.

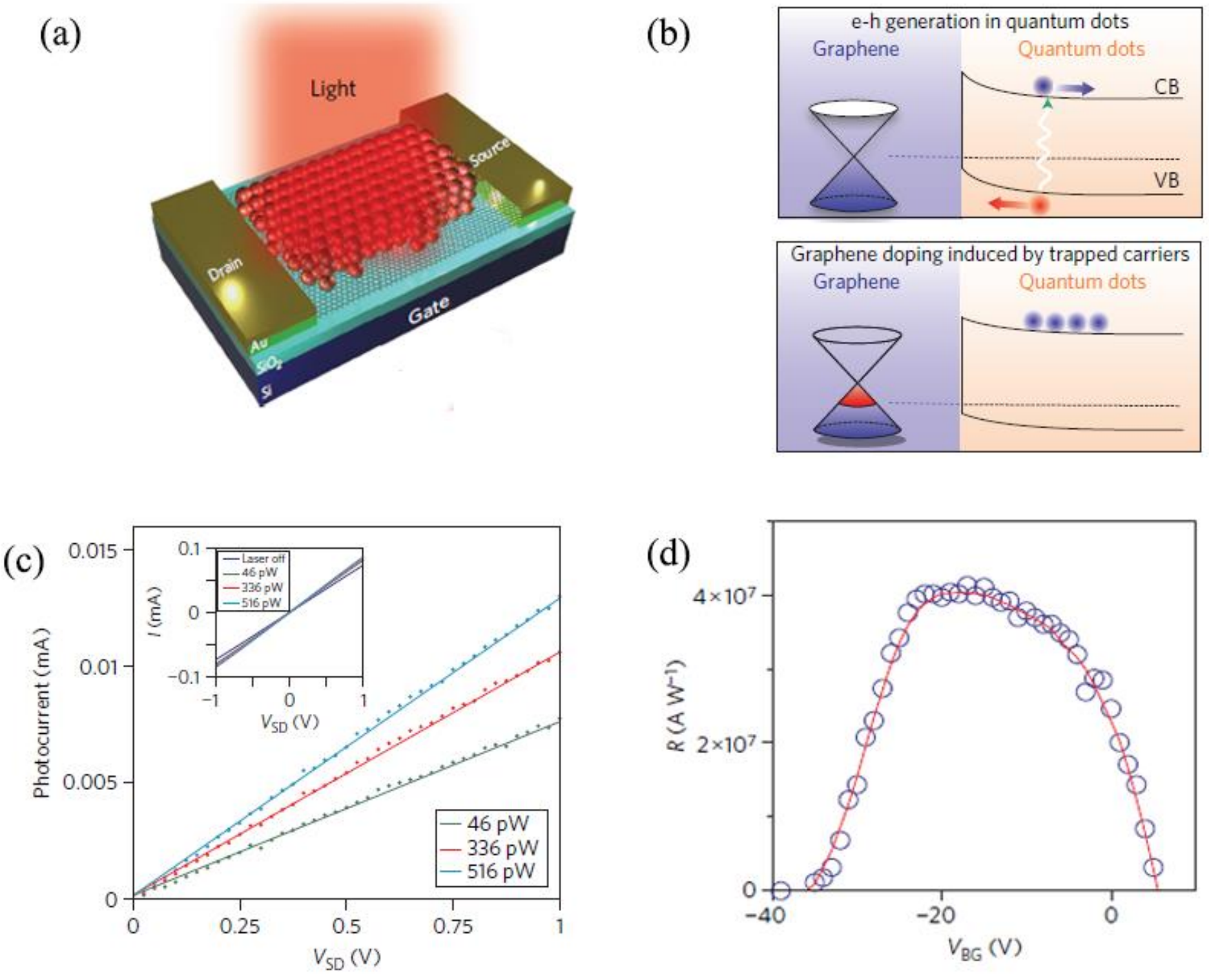

**Figure 1** (a) The schematic diagram of a hybrid graphene-PbS quantum dots phototransistor. (b) The energy band diagram at the graphene/PbS QDs interface. (c) The linear dependence of the photocurrent on the applied drain-source voltage for different optical powers (46, 336, 516 pW). $V_{BG}$=0. Inset: The total current ($I=I_{dark}+I_{photo}$) as a function of the applied drain-source voltage for different optical powers. (d) The responsivity as a function of the applied back-gate voltage $V_{BG}$. $V_{SD}$=5 V. (a-d) are reproduced from Ref.[32] with permission from the Nature Publishing Group.

Another important two-dimensional material, transition metal dichalcogenides

(TMDs), has also been extensively studied in recent years.[43-46] However, due to the bandgap limitation and ultrathin thickness, the response spectral range and responsivity of TMDs-based infrared photodetectors are limited. Similarly, in order to enhance light absorption and broaden the response spectral range, a strategy for integrating TMDs materials with quantum dots has also been exploited.[31, 35, 36] Huo et al. demonstrated $MoS_2$-HgTe QDs hybrid photodetectors beyond 2μm.[31] The infrared detection capability of this detector is mainly caused by HgTe QDs thin film, which acts as a sensitizing layer, and the response spectrum range can be further extended to the mid-wave infrared and long-wave infrared ranges by adjusting the size of HgTe quantum dots.[47, 48] A $TiO_2$ buffer layer between the $MoS_2$ transistor channel and the HgTe QDs sensitizing layer was used as a protective layer of the $MoS_2$ channel and as an n-type electron acceptor medium to form an efficient *p–n* junction with the HgTe QDs at the interface facilitating the charge transfer to $MoS_2$ channel.[31, 49] Thus, the dark current can be adjusted by the gate voltage and the specific detectivity D* (~$10^{12}$ Jones) at a wavelength of 2 μm can be obtained. The detector can operate at room temperature and exhibit sub-milliseconds response and high responsivity of ~$10^6$ A/W, which demonstrates that the hybrid 2D-QDs structure is an effective method to improve the performance of photodetector based on 2D materials.

In order to obtain high gain to make up for the weak light absorption of 2D materials, many infrared photodetectors based on hybrid 2D-QDs heterostructures are phototransistors that have an electronic passive sensitization layer composed of strong light absorbing material.[32, 50, 51] Thus, the external quantum efficiency (EQE), response speed and linear dynamic range are limited by thin sensitizing layers thickness, long trapping times, and high density of the sensitizing centers trap-state, respectively.[32, 41, 52] In order to solve these problems, Nikitskiy et al. proposed a photodetector that integrates an electrically active colloidal QD photodiode with a graphene transistor.[34] The device is composed of a bottom graphene channel, a 300-nm-thick PbS QDs layer and a top ITO electrode. The photodiode operation and phototransistor operation of the hybrid 2D-QDs photodetector are shown in Figure 2a and b, respectively. Photogenerated electron-hole pairs are created at the QDs layer, and then drifted by the built-in electric field at the interface and by the applied bias voltage to form photocurrent directly in photodiode operation. However, after the electron-hole pairs are separated, electrons remain in the QDs layer and holes are driven by depletion region on the graphene/PbS QD interface and transferred to graphene in phototransistor operation. For $V_{TD}>0$, the width of the depletion region expands, increasing the efficient charge collection area, and thus the external quantum efficiency also increases. The graphene-QDs photodiode responsivity and EQE are shown in Figure 2c. With increasing $V_{TD}$, EQE increased from 10% at $V_{TD}=0$ V to 75% at $V_{TD}=1.2$ V (limited by reflection), indicating the importance of $V_{TD}$ for improving EQE. Figure 2d shows the EQE and responsivity in phototransistor operation, agreeing with the dependence of $EQE_{TD}$ on $V_{TD}$ in photodiode operation, which demonstrates that the EQE of this integrated configuration is determined by the photodiode. Moreover, the responsivity is $5\times10^6$ V/W at $V_{TD}=1.2$ V, indicating that the phototransistor still has a high gain of $10^5$ compared with the photodiode operation. With increasing $V_{TD}$, the temporal response accelerates and the linear dynamic range extends. The electrical 3 dB bandwidth of the detector reaches 1.5 kHz for the optimum $V_{TD}$, and the linear dynamic range extends from 75 dB at $V_{TD}$=0 V to 110 dB at $V_{TD}$=1.2 V. The above results prove that the design and optimization of infrared photodetectors based on hybrid 2D-QDs heterostructures can effectively enhance the detection performance.

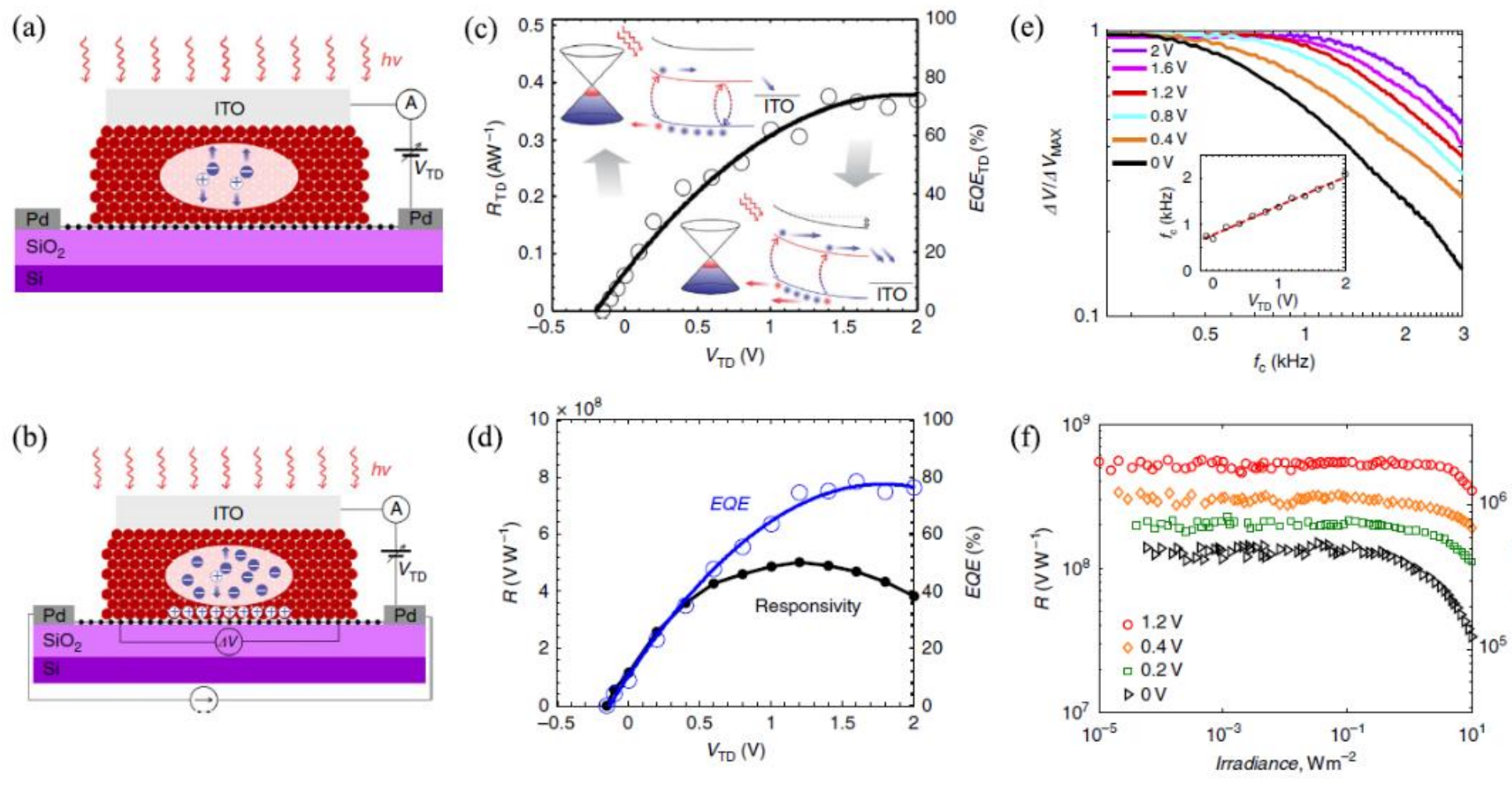

**Figure 2** (a-b) Schematic of photodiode operation and phototransistor operation of the hybrid graphene-PbS QDs photodetector, respectively. (c) Responsivity and external quantum efficiency EQE of photodetector in photodiode operation as a function of $V_{TD}$. Inset: energy band diagrams of device with and without $V_{TD}$ bias. (d) Responsivity and EQE of photodetector in phototransistor operation as a function of $V_{TD}$. (e) Normalized photoresponse versus light modulation frequency. The specified $V_{TD}$ values are shown in the legend. Inset: The linear dependence of extracted 3 dB bandwidth values on applied $V_{TD}$. (f) Responsivity of the detector as a function of incident irradiance in the range of $10^{-5}$ to 10 $Wm^{-2}$, which shows a significant extension of the linear dynamic range with increasing $V_{TD}$. (a-f) are reproduced from Ref.[34] with permission from the Nature Publishing Group.

### 2.2 Nanowires combined with 2D materials

One-dimensional nanomaterials, such as single-wall carbon nanotubes (SWNTs), have unique and excellent optical and electrical properties due to 1D quantum confinement effect.[53] Because of the type of tube chiralities, metallic or semiconducting properties are exhibited, and the carrier mobility of semiconducting SWNTs can reach up to $10^5$ $cm^2V^{-1}s^{-1}$ at room temperature.[54-56] SWNTs have broadband and tunable absorption, so it has great potential in the field of infrared detection. However, the photoresponsivity is limited by the small light-absorbing area, and the responsivity of single-tube photovoltaic devices is always $<10^{-3}$ A/W.[57] Paulus et al. reported that the junction between a metallic SWNT and graphene can perform effective charge transfer.[58] This is because that the SWNT is able to dope the graphene in a spatially controlled way without affecting its $sp^2$ lattice structure.[58] Liu et al. demonstrated a high-performance photodetector based on a planar atomically thin SWNT-graphene hybrid film across visible to near-infrared range (400-1,550 nm).[50] Schematic of the photodetector and atomic force microscope (AFM) characterization of SWNT-graphene hybrid film are shown in Figure 3a. SWNTs have a diameter in the range of 1.0-1.6 nm, and the $S_{11}$ and $S_{22}$ bands are found to be located at ~1,800 and ~1,000 nm with higher SWNTs loadings. The photogenerated electron-hole pairs are generated at the SWNT/ graphene interface under optical illumination, and then separated by the built-in electric field. Electrons are transferred from SWNTs to graphene, while holes are trapped in the SWNTs, forming a photogating effect. Figure 3b shows that the source-drain current ($I_{SD}$) as a function of the applied back-gate voltage ($V_G$). The Dirac point shifts toward the lower $V_G$ as the light increases, indicating that the graphene resistance can be effectively modulated by the photogating effect. Unlike other photodetectors that use the photogating effect to increase the responsivity, which generally has slow response speed, the photodetector based on hybrid SWNTs-graphene exhibits a fast response time of ~100 μs (Figure 3c). This is mainly due to the high mobility of graphene and SWNTs and fast charge transfer at the SWNTs-graphene interface, and the field-effect mobility of electrons and holes are measured as 3,920 and 3,663 $cm^2V^{-1}s^{-1}$, respectively. The transit time of the hybrid SWNTs-graphene photodetector is estimated to be ~ $10^{-9}$s, so the photoconductive gain is ~ $10^5$ and the gain-bandwidth product is ~$10^9$ Hz. Moreover, since SWNTs act as a sensitizing layer, the detector also exhibits the characteristics of a broadband photoresponse. The relationship between the responsivity and the incident optical power at different illumination wavelengths in the range of visible to near-infrared range (405, 532, 650, 980 and 1,550 nm) is shown in Figure 3d, which shows a high responsivity of >100 A/W. Further, Liu et al. fabricated a photodetector based on graphene/carbon nanotube hybrid films on a PET substrate and measured the performance under bending conditions.[59] It was found that this device has good robustness against repetitive bending, which fully demonstrated the huge application potential of this heterostructure in the field of flexible electronics.

Jariwala et al. demonstrated a gate-tunable carbon nanotube-$MoS_2$ heterjunction p-n diode.[60] Since both monolayer $MoS_2$ and semiconducting SWNTs have direct bandgaps and tightly bound excitation states in absorption spectra,[61] the photodiode based on hybrid $MoS_2$-SWNTs also exhibits excellent detection performance. The external quantum efficiency (EQE) and the response time are ~25% and <15 μs, respectively. Figure 3e shows that photocurrent spectral response can be tuned by changing the gate voltage. The photocurrent in the near-infrared region continuously decreases as the gate voltage decreases. Furthermore, the performance of devices based on 1D-2D heterostructures can be further optimized by strain modulation. For example, the responsivity of the photodetector based on the graphene-ZnO nanowire heterostructure is improved by 26% under 0.44% tensile strain on the ZnO nanowire.[62] The above results show that it is significant to carry out the research of the hybrid 1D-2D heterostructure in the field of infrared detection.

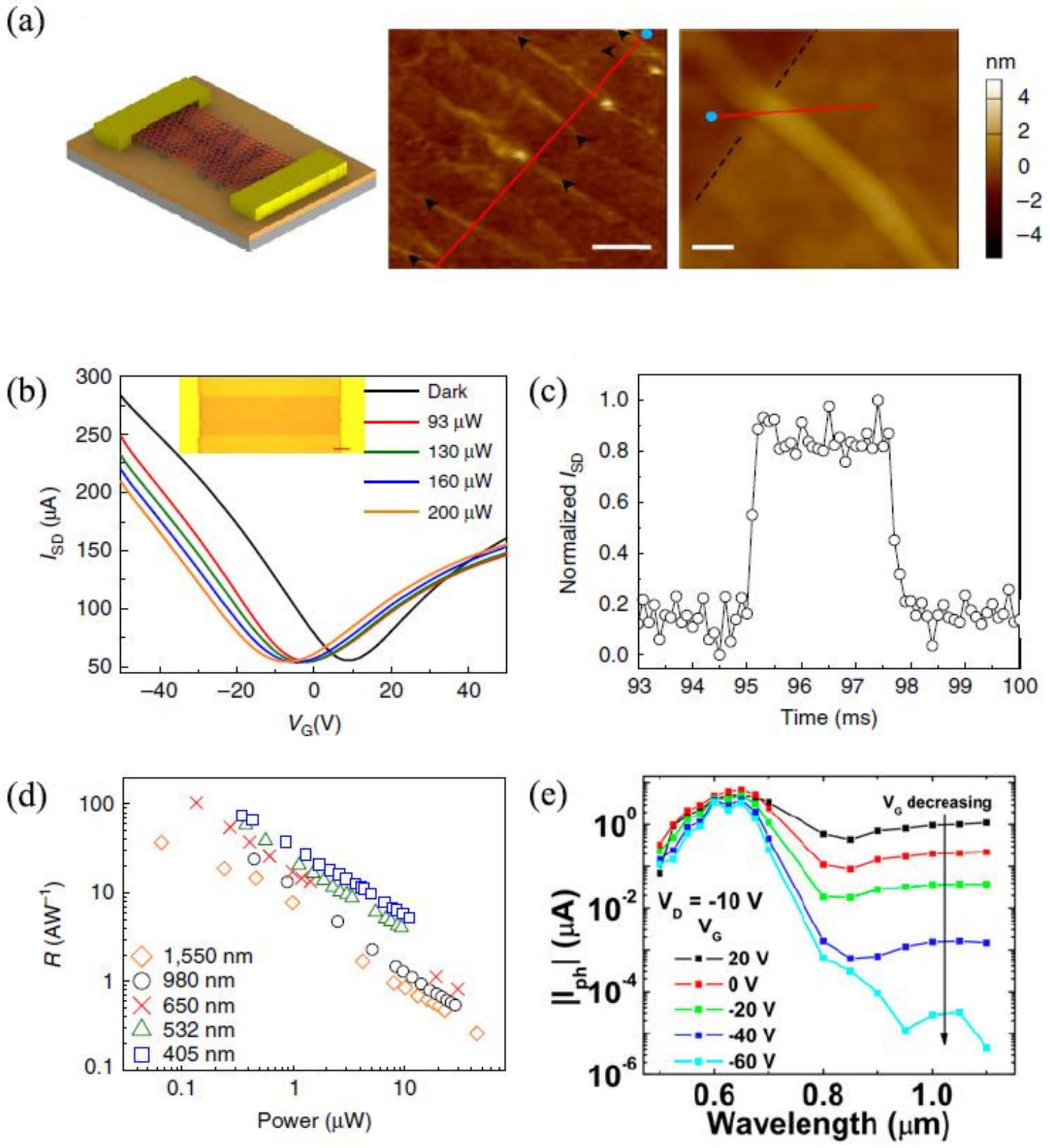

**Figure 3** (a) Schematic of the photodetector and AFM characterization of SWNT–graphene hybrid film (middle: scale bar, 200 nm; right: scale bar, 20 nm). (b) The $I_{SD}$-$V_G$ transfer curve of device with increasing 650 nm illumination powers. $V_{SD}$=0.5 V. Inset: the optical image of the SWNT-graphene device (Scale bar, 10μm). (c) Temporal response of the SWNT–graphene hybrid photodetector under 650 nm illumination. (d) The dependence of responsivities of photodetector on the optical power at different illumination wavelengths from visible to near-infrared range (405, 532, 650, 980 and 1,550 nm). (e) Photocurrent spectral response can be tuned by varying the gate voltage. $V_D$=-10 V. (a-d) are reproduced from Ref.[50] with permission from the Nature

Publishing Group. (e) are reproduced from Ref.[60] with permission from the National Academy of Sciences.

## 2.3 2D vertical and lateral heterostructures

### 2D vertical heterostructures

Two-dimensional vertical heterostructure is fabricated by stacking different two-dimensional materials in the vertical direction. This heterostructure is generally prepared by targeted transfer method or CVD growth technique.[63-68] 2D vertical heterostructures have many advantages, such as atomically sharp interface, no atomic diffusion between layers, and no strict lattice matching problems due to weak van der Waals interaction between 2D materials. Moreover, the strong interlayer coupling and ultrafast transfer of charge carriers make it have great potential in the field of optoelectronics and electronic devices.[68-73]

In order to achieve infrared detecting, photodetectors must have a broad spectral response. However, the development of TMDs for infrared detection is hindered due to the lack of 2D materials with narrow bandgap. Monolayer $\alpha$-$MoTe_2$ has a band gap of 1.1 eV and can absorb light in the near-infrared range.[74, 75] The response spectral of the photodetectors based on the $\alpha$-$MoTe_2$/$MoS_2$ dichalcogenide heterojunction covers in the range from the visible to near-infrared (800 nm).[76] However, this is not enough for mid-infrared and far-infrared detection, and the response range of the photodetector should be further extended. Graphene can absorb light indiscriminately due to the gapless band structure and monolayer black phosphorene shows a direct optical band gap of 1.3 eV,[77, 78] so they are effective absorption material in the infrared region.

Long et al. demonstrated a broadband photodetector based on $MoS_2$-graphene-$WSe_2$ heterostructure covering visible to short-wavelength infrared range at room temperature.[79] The schematic diagram and optical image of the photodetector based on a $MoS_2$-graphene-$WSe_2$ heterostructure are shown in Figure 4a. The p-doped $WSe_2$ is induced by large work-function metal Pd, whereas $MoS_2$ flakes remain n-type due to Fermi level pinning,[80] and thus forming a p-n junction with a strong internal electric field of ~$2\times10^8$ V/m. Photogenerated electron-hole pairs can be effectively separated in the presence of the internal electric field. Therefore, the dark current of the device is effectively suppressed and the specific detectivity increases accordingly. Figure 4b shows the responsivity R and specific detectivity D* of the photodetector for wavelengths ranging from 400 to 2,400 nm with an interval of 100 nm. R and D* are as high as $10^4$ A/W and $10^{15}$ Jones in the visible region, while they decrease sharply to 100 mA/W and $10^9$ Jones in the infrared wavelength of 2400 nm, respectively. Figure 4c explains the reason for the sharp changes in responsivity and specific detectivity. The band gap of $MoS_2$ ($E_{g1}$) and $WSe_2$ ($E_{g2}$) are 1.88 eV and 1.65 eV, respectively, and the corresponding wavelengths are 660 nm and 750 nm.[21, 29, 81] When the incident photon energy hν exceeds the bandgap of the material, the material can absorb photons to generate electron-hole pairs. Therefore, for $h\nu > E_{g1} > E_{g2}$, $MoS_2$, graphene and $WSe_2$ all can generate photoelectron electron-hole pairs, resulting in a higher responsivity. For $h\nu < E_{g1} < E_{g2}$, only graphene can generate photoelectron electron-hole pairs, and thus the photocurrent is drastically reduced. Qiao et al. reported a broadband photodetectors based on graphene-$Bi_2Te_3$ heterostructure (Figure 4d).[82] $Bi_2Te_3$ has only 2.7% lattice mismatch with graphene, so the graphene-$Bi_2Te_3$ heterostructure can be directly prepared on graphene by Van der Waals epitaxial growth method.[83] Figure 4e plots photocurrent of devices based on pure graphene and graphene-$Bi_2Te_3$ heterostructure as a function of source-drain voltage without applied gate bias, which shows that the latter is ten times higher than the former. Both graphene and $Bi_2Te_3$ absorb light to generate photoelectron-hole pairs, which are subsequently separated by a built-in electric field. Then, photogenerated electrons in graphene are transferred into $Bi_2Te_3$, while the photogenerated holes in $Bi_2Te_3$ are transferred into graphene, which suppresses the recombination of carriers and generates a larger photocurrent. Figure 4f shows the dependence of photocurrent on wavelengths ranging from 400 nm to 1,550 nm. Due to the gapless nature of graphene and a small bandgap (0.15-0.3 eV) of $Bi_2Te_3$,[84] both graphene and $Bi_2Te_3$ are effective absorbent materials for infrared light, and thus the detector exhibits broadband detection characteristics. The effective detection range can be extended to the near infrared (980 nm) and c-communication band (1,550nm). The responsivity is 35 A/W, 10 A/W and 0.22 A/W in the visible (532 nm), infrared (980 nm) and c-communication band (1,550 nm) regions, respectively.

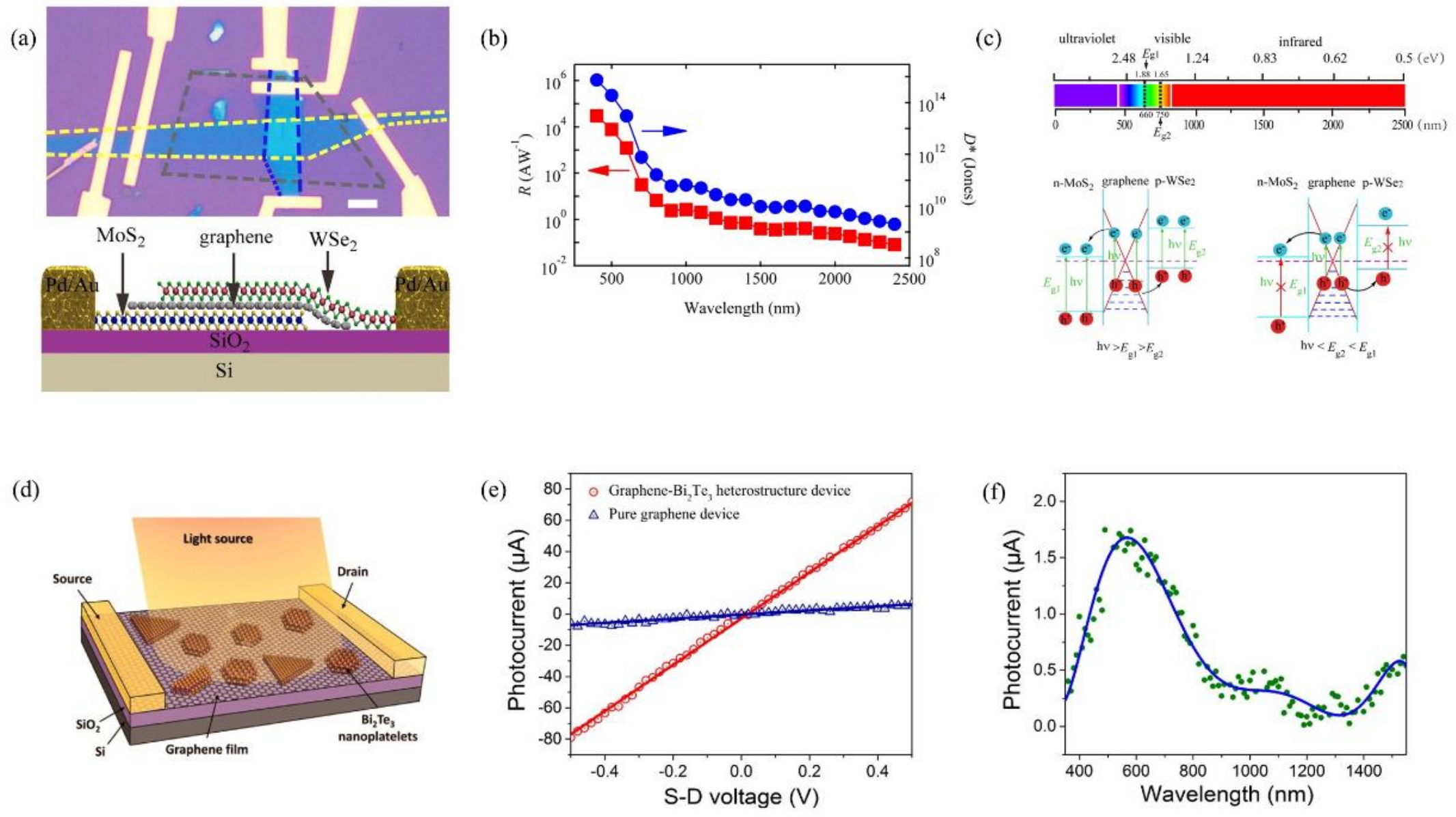

**Figure 4** (a) Top: optical image of the photodetector based on a $MoS_2$−graphene−$WSe_2$ heterostructure. Scale bar, 5 μm. Bottom: Schematic diagram of the device. (b) Responsivity and specific detectivity of the photodetector for wavelengths ranging from 400 to 2,400 nm with an interval of 100 nm. (c) Top: Spectra from ultraviolet to infrared. The band gap of $MoS_2$ ($E_{g1}$) and $WSe_2$ ($E_{g2}$) as well as corresponding wavelengths are plotted on the graph, respectively. Bottom left: For hν >$E_{g1}$>$E_{g2}$, $MoS_2$, graphene and $WSe_2$ all can generate photogenerated electron-hole pairs. Bottom right: For hν <$E_{g1}$<$E_{g2}$, only graphene can generate photogenerated electron-hole pairs. (d) The schematic diagram of the device based on graphene-$Bi_2Te_3$ heterostructure. (e) The dependence of photocurrent of devices based on pure graphene and graphene-$Bi_2Te_3$ heterostructure on source-drain voltage $V_{SD}$. $V_G$=0 V. (f) Dependence of photocurrent on wavelengths ranging from 400 nm to 1,550 nm. (a-c) are reproduced from Ref.[79] with permission from the American Chemical Society.

(d-f) are reproduced from Ref.[82] with permission from the American Chemical Society.

So far, highly polarization sensitive infrared photodetector based on two-dimensional materials working at room temperature are still lacking. Ye et al. first demonstrated a broadband photodetector based on a vertical photogate heterostructure of BP-on-$WSe_2$ with highly polarization sensitive for infrared detection (Figure 5a).[85] $WSe_2$ acts as a conductive channel, while BP with a small direct bandgap and highly anisotropic crystal structure acts as a photogate.[18, 19, 77] Both $WSe_2$ and BP absorb light efficiently and generate photogenerated electron-hole pairs under visible illumination, which are then separated by the built-in electric field. The photo-generated electrons in BP are transferred to $WSe_2$ and the photo-generated holes in $WSe_2$ are transferred to BP, which effectively reduces the recombination efficiency. An increasing number of photo-generated electrons in $WSe_2$ are driven by source-drain bias to form photocurrent, thereby enhancing the responsivity. However, due to the band gap limitation of $WSe_2$, only BP with a small band gap is an effective absorption material to generate photo-generated free carriers under infrared illumination. So the photoresponse in the infrared region will be smaller than in the visible region. The responsivity is ~$10^3$ A/W under visible illumination (637 nm), while the responsivity is ~0.5 A/W under infrared illumination (1,550 nm). Figure 5b shows the photoconductive gain G and detectivity D* of the photodetector for wavelengths ranging from 400 to 1,600 nm. The detectivity in the visible and infrared regions are $10^{14}$ and $10^{10}$ Jones, and the gain in the visible and infrared regions are $10^6$ and $10^2$, respectively, demonstrating excellent broadband detection performance. The optical image of the BP-on-$WSe_2$ photodetector and the relationship between the photocurrent image of the photodetector at 1,550 nm and the polarization angle are shown in Figure 5c. The advantage of this device structure is that it can eliminate the disturbance of two-fold polarization-dependent photocurrent resulted from the geometric edge effect at the metal-BP edge and fully collect the photo-generated free carriers isotropically for highly polarization-sensitive infrared detection.[85] As can be seen from the photocurrent image, the polarization photoresponse originates from the overlap of BP and $WSe_2$, and it changes obviously with the polarization angle changes. The polar plot in Figure 5d shows the relationship between the photocurrent and the polarization angle of the incident light. The maximum photocurrent, corresponding responsivity of ~40 mA/W, is taken along the horizontal axis (defined as $0^o$ polarization), while the minimum photocurrent, corresponding responsivity of ~6.8 mA/W, is taken along the vertical axis (defined as $90^o$ polarization).[85] The ratio of the maximum responsivity to the minimum responsivity is about 6, showing high polarization sensitive infrared detection performance. Similarly, the mid-wave infrared photodetector based on BP and $MoS_2$ heterostructure designed by Bullock et al. also utilized the anisotropy of BP to achieve polarization detection, which is bias-selectable and operates without the need for external optics.[86] The device exhibits high EQE (35%) and D* ($1.1\times10^{10}$ cm $Hz^{1/2}$ $W^{-1}$) at room temperature.

The above results show that the 2D vertical heterostructure can effectively extend the response range of the photodetector to achieve infrared detection. In addition, CVD epitaxial growth technology provides a guarantee for the scaled preparation of vertical heterostructures. By combining 2D materials with different unique and excellent physical properties, a multifunctional infrared detector with excellent detection performance can be prepared.

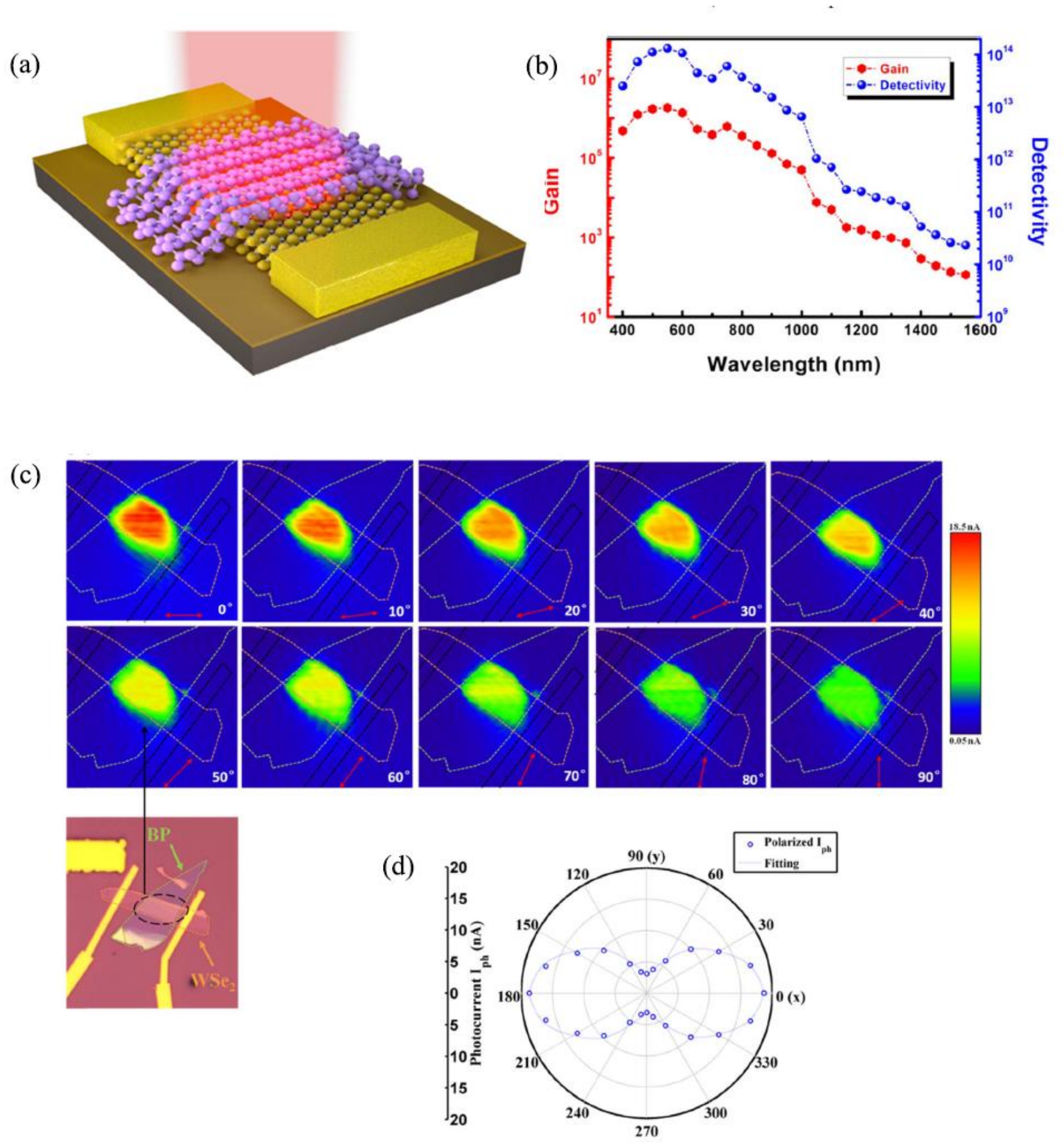

**Figure 5** (a) Schematic diagram of the photodetector with BP-on-$WSe_2$ photogate structure. (b) The photoconductive gain G and detectivity D* of the photodetector for wavelengths ranging from 400 to 1,600 nm at 1 $mW/cm^2$ incident light power density.

$V_{ds}$=0.5 V. (c) Top: the photocurrent image of the photodetector under 1,550 nm illumination with different light polarizations (red arrows). Bottom left: the optical image of the photodetector with BP-on-$WSe_2$ photogate structure. (d) The photocurrent as a function of light polarization at 100 mW/cm$^2$ incident light power density. $V_{ds}$=0.5 V. (a-d) are reproduced from Ref.[85] with permission from Elsevier.

**2D lateral heterostructures**

The heterostructures mentioned above are basically vertical, and the built-in electric field at the interface promotes the separation of photogenerated electron-hole pairs and charge transfer, thereby reducing the recombination and accelerating the response speed. In addition, two-dimensional lateral heterostructures, a heterostructure composed of materials with different work functions or homojunction composed of the same material but different doping types in the lateral direction, form a built-in electric field at the interface and play a similar role. At present, the methods for forming lateral heterostructures generally include chemical doping,[87, 88] electrostatic doping,[89-91] and one-pot or two-step chemical vapor deposition (CVD) growth methods.[64, 65, 92-96]

Recently, Yu et al. demonstrated a lateral black phosphorene P–N junction formed via chemical doping for high performance near-infrared photodetector.[88] The n-type doping of few-layer black phosphorene is achieved by using benzyl viologen (BV) as an electron dopant, which can obtain a high electron concentration while still maintain the high carrier mobility of BP as the processes do not induce defects in the crystal lattices, and $Al_2O_3$ is used to selectively protect part of the area of a p-type BP at the same time.[88] The electron doping concentration of BP can be adjusted by doping time, and the built-in electric field at the P-N junction can effectively separate photogenerated electron-hole pairs and reduce recombination efficiency. Photocurrent tests show that the off-current under 1.47 μm illumination is 4 orders of magnitude higher than in dark condition, demonstrating that the photocurrent dominates over thermionic and tunneling currents in the photodiode operation.[88] The photodetector exhibits high responsivity of ~180 mA/W, fast response of tens of milliseconds and high detectivity in the range of $10^{11}$-$10^{13}$ Jones under infrared light illumination ($\lambda$=1.47 μm).

However, the current methods for preparing lateral heterojunctions have their own limitations. Chemical doping has the characteristic that it is difficult to find suitable dopants. The density of the carrier modulated by the electrostatic gating is relatively low, so the modulation effect is limited.[19, 97] CVD technology is an effective method for large-scale preparation of lateral heterostructures, but the main focus is on epitaxial growth of TMDs with large band gap at present, and thus CVD technology has limitation in the field of infrared detection. The infrared light detection technology based on 2D lateral heterostructure needs further research and development.

# 3. On-chip structure

## 3.1 2D materials on waveguide

Two-dimensional materials show strong interaction with light, but the spectral response range and responsivity of the photodetector based on 2D materials are limited due to the reason of the atomically ultra-thin thickness and the limitation of the band gap. In order to obtain better infrared detection performance to achieve practical applications, it is necessary to enhance the infrared light absorption of two-dimensional materials. Integrating two-dimensional materials with waveguides is a good strategy to enhance light absorption.[91, 98-104] The propagation of light in the waveguide is accompanied by the evanescent field, which can be absorbed by the two-dimensional materials to generate photogenerated electron-hole pairs.[105-108] Different with the incident light perpendicularly incident on the two-dimensional materials, the in-plane absorption of two-dimensional materials makes its nanoscale thickness to be no longer a major limiting factor for light absorption. The length of the two-dimensional material in contact with the waveguide is on the order of micrometers, so the length of the interaction with the light is also on the order of micrometers and therefore can effectively enhance light absorption.

Recently, a series of similar but independent waveguide integrated graphene photodetectors have been reported,[98, 101, 102] and the device structures are shown in Figures 6a, c and e, respectively. Figure 6a shows a scanning electron microscope (SEM) image of a waveguide-integrated graphene photodetector using a GND-S-GND configuration. It should be noted that in addition to the length L of graphene, the width W of central electrode S in this device also affects the light absorption of the graphene. The absorption coefficient of metal electrode $\alpha_M$ increases with increasing electrode width W. For W>100 nm (W>160 nm), the light absorption of the metal electrode will exceed the absorption of monolayer (bilayer) graphene. However, the contact resistance between metal and graphene also increases with decreasing width of electrode, so the metal electrode absorption coefficient and contact resistance should be considered comprehensively to obtain the best device performance. Figure 6b shows the absorption of graphene as a function of graphene length L at different widths, indicating that as long as the appropriate W and L values are obtained, the light absorption of graphene can be greatly enhanced. For example, when W = 100 nm, more than 50% light absorption can be obtained with only 22 μm length bilayer graphene. Figure 6d shows an SEM image of a waveguide-integrated graphene photodetector with high responsivity and photocurrent image at zero bias. A 53 μm bilayer graphene was overlaid on the waveguide, and two electrodes with waveguide distances of 100 nm and 3.5 μm, respectively, were used to collect the photogenerated carriers. The lateral metal-doped junction overlapping the waveguide is formed by doping of metal electrode 100 nm away from the waveguide, and effectively promotes the separation of photogenerated electron-hole pairs. The photocurrent image shows that the built-in electric field between metal-doped graphene and graphene appears at the metal and graphene interface and leads to two narrow photocurrent regions. Moreover, the region of maximum photocurrent coincides with the position of waveguide. Wavelength transmission measurements show that the transmission loss are 0.1 dB and 4.8 dB before and after the graphene transfer, respectively, indicating that the optical absorption of the waveguide-integrated graphene is greatly increased. The photodetectors demonstrated an ultrafast response rates exceeding 20 GHz at zero bias and a high response rate of 0.108 A/W at 1 V bias. Thanks to the gapless graphene, the spectral response of photodetector ranged from 1,450 nm to 1,590 nm, which can effectively detect the c-communication band of the near-infrared region. The response spectrum of the waveguide-integrated graphene photodetector can be further extended to the mid-infrared region. Xu and coworkers demonstrated a graphene/silicon heterostructure waveguide photodetector with a responsivity of up to 0.13 A/W at room temperature in the mid-IR region (2.75 μm).[102] In addition, this photodetector exhibits a low dark current due to the presence of a graphene/silicon heterojunction potential barrier, and the on/off photocurrent ratio is 30 at a bias of -1.5 V for 2.75 μm illumination at room temperature (Figure 6f). This work further explains the photocurrent generation mechanism of waveguide-integrated graphene in the visible light region, near infrared region and mid-infrared region. For the visible light region, since silicon has an indirect bandgap of 1.1 eV, photoresponse is mainly due to electron-hole pairs generated after the silicon absorbs light and graphene acts as a photogenerated free carriers conduction channel in this process. For the near-infrared region (1.55 μm), the photon energy is more than twice that of the graphene Fermi energy level and the infrared light cannot be absorbed by silicon. The photoexcitation of graphene takes place in a direct transition way that electrons are excited from the valence band to the conduction band. For the mid-infrared region (2.75 μm), the photon energy is less than twice the energy difference between the Fermi level and the Dirac point, thus preventing direct transitions of graphene.[102] The excitation of graphene follows the indirect transition processes, and electrons are excited from Fermi level in the valence band to the conduction band with a large momentum mismatch under mid-infrared light illumination.

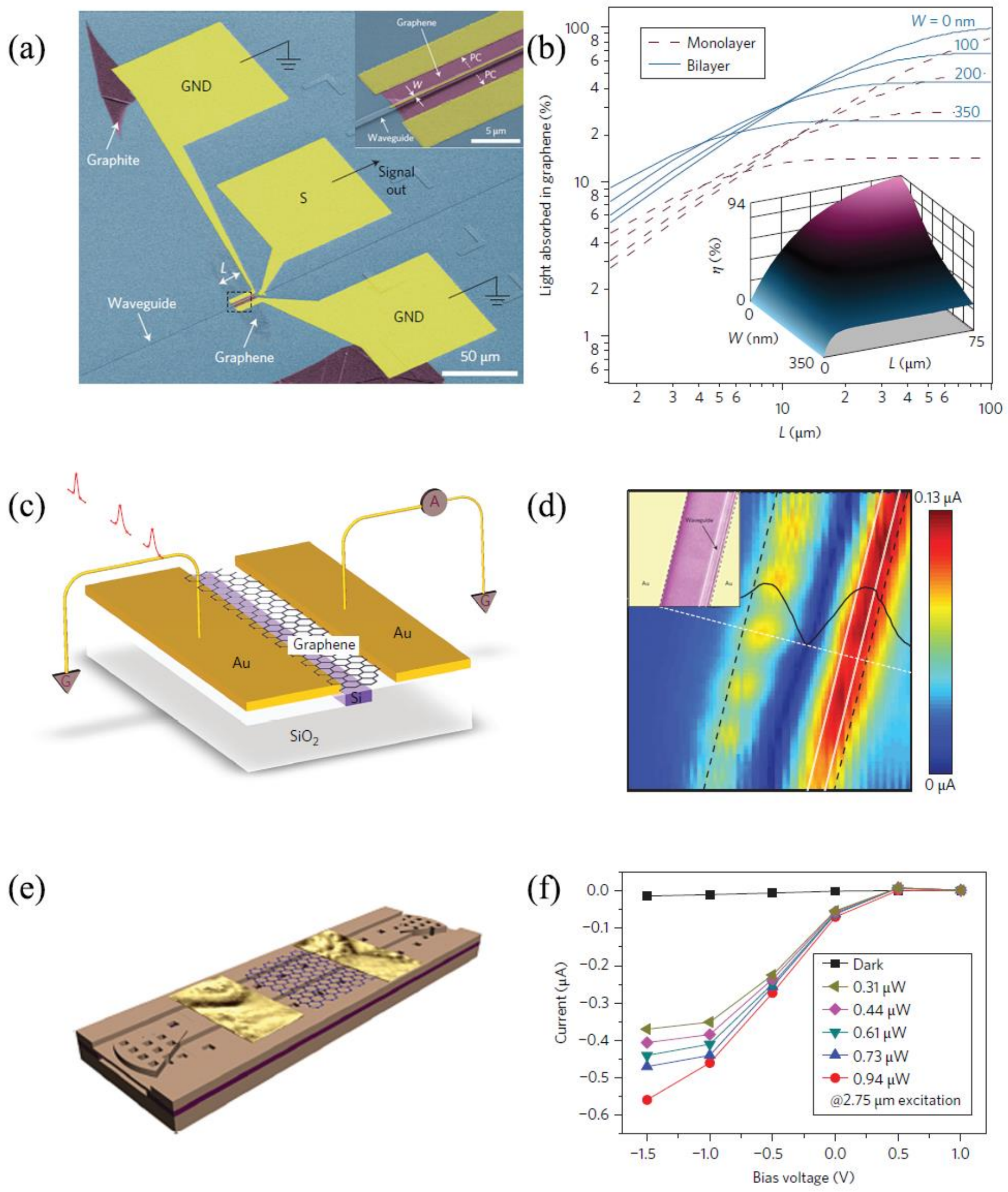

**Figure 6** (a) SEM image of a waveguide-integrated graphene photodetector using a GND-S-GND configuration (false colour). Inset: An enlarged view of the inside of the black dashed line. (b) The absorption of graphene as a function of graphene length L at different widths W. Solid line: bilayer graphene. Dashed line: monolayer graphene. (c)

Schematic of the chip-integrated graphene photodetector. (d) The spatially resolved photocurrent image of the waveguide-integrated graphene photodetector at zero bias. The black solid line shows the relative potential distribution across the graphene channel along the white dashed line. Inset: corresponding SEM image. (e) Schematic of a graphene/silicon heterostructure waveguide photodetector. (f) Photocurrent as a function of applied bias voltage with 2.75 μm laser illumination at different light power. (a-b) are reproduced from Ref.[101] with permission from the Nature Publishing Group. (c-d) are reproduced from Ref.[98] with permission from the Nature Publishing Group. (e-f) are reproduced from Ref.[102] with permission from the Nature Publishing Group.

Similarly, Youngblood et al. demonstrated a waveguide-integrated few-layer black phosphorus photodetector with high responsivity and low dark current.[103] Figure 7a shows the structure of the photodetector, having a graphene top-gate, a BP channel and the waveguide patterned on the silicon-on-insulator substrate. Since the incident light propagates along the waveguide, the optical interaction length is equal to the width of the black phosphorus covering on the waveguide (6.5 μm). The device can absorb 78.7% of the optical power in the waveguide, of which only 17.5% is absorbed by the black phosphorus and the rest is absorbed by the graphene top-gate.[103] Figure 7c shows that the photocurrent versus the gate voltage $V_G$ and the bias voltage $V_{DS}$. $V_G$ can electrostatically tune the hole doping concentration of black phosphorus and the BP is nearly intrinsic at $V_G$=-8 V. When black phosphorus is in a low doping state (-10 V < $V_G$ < -1 V), the photocurrent is larger and its sign is positive relative to $V_{DS}$, which suggests photocurrent generation mechanism is photovoltaic effect at this time.[109-111]

For $V_G$ > 0 V, BP is tuned to be more heavily n-doped, the photocurrent is getting smaller and its sign changes to negative relative to $V_{DS}$, and the photocurrent generation mechanism is bolometric effect at this time.[109, 112] The frequency response of the photocurrent shown in Figure 7b can also demonstrate the photocurrent generation mechanism at different gate voltages. The roll-off frequency is measured to be 2.8 GHz when BP is in a low doping state ($V_G$=-8 V), showing a fast response due to photovoltaic effect. But the roll-off frequency is measured to be 0.2 MHz when BP is in a high doping state ($V_G$=8 V), showing a slower response due to bolometric effect. Figure 7d shows the intrinsic responsivity R and internal quantum efficiency IQE as a function of bias for different thicknesses. The intrinsic responsivity and internal quantum efficiency of 11.5-nm-thick BP device with top-gate are 135 mA/W and 10% at $V_{bias}$=-0.4 V, respectively. For 100-nm-thick BP device without top-gate, a larger bias voltage can be applied because there is no need to consider the breakdown of the gate insulation layer. Therefore, this photodetector shows better detection performance with high intrinsic responsivity (657 mA/W) and high internal quantum efficiency (50%) at $V_{bias}$=-0.4 V. Unlike gapless graphene, BP has a finite direct band gap,[113] which make the dark current of waveguide-integrated BP photodetector extreme low (220 nA) at $V_G$=-8 V and $V_{bias}$=-0.4 V.

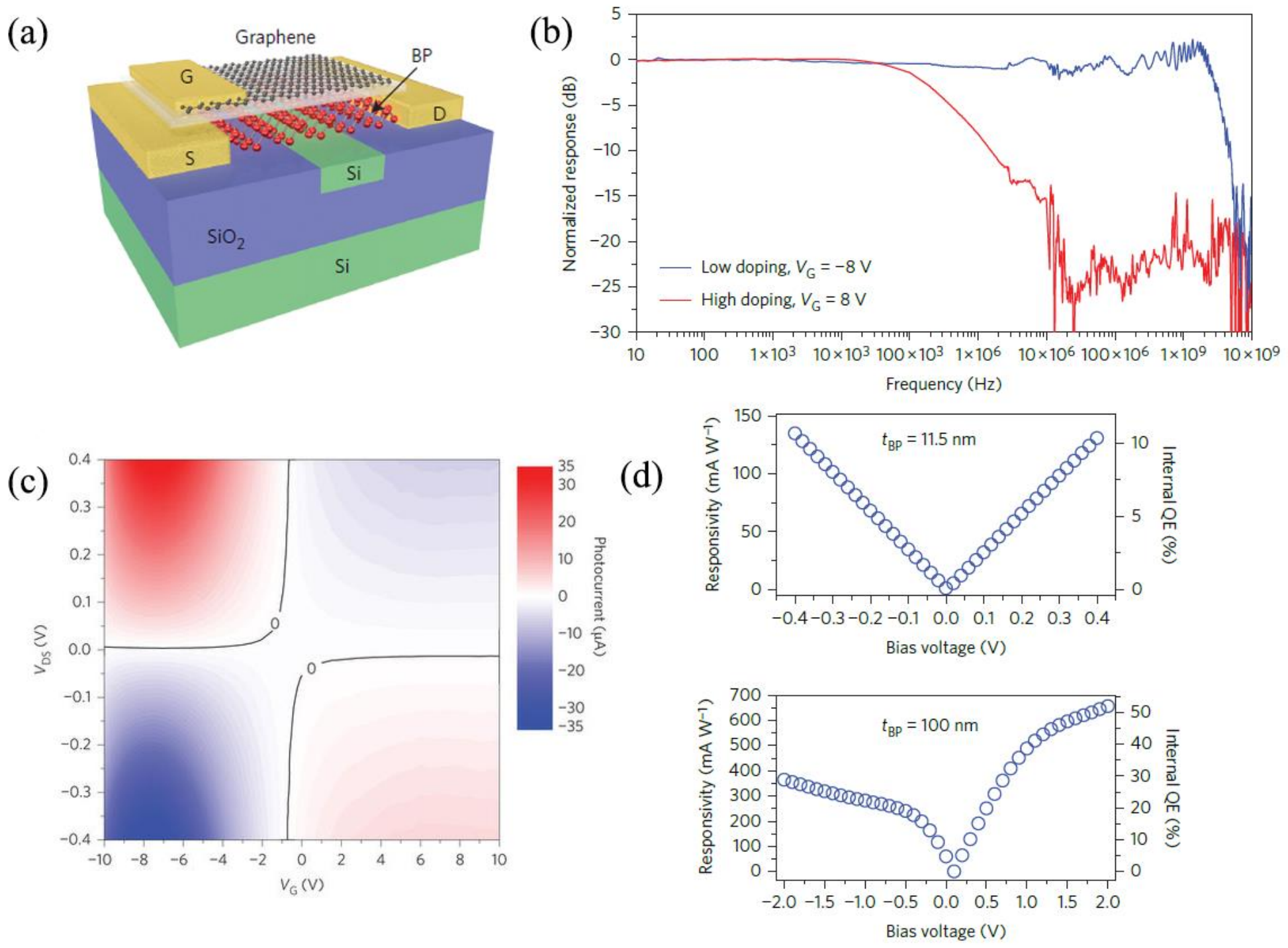

**Figure 7** (a) Schematic of the waveguide-integrated few-layer BP photodetector. (b) The frequency response of the waveguide-integrated BP photodetector at low doping state ($V_G$=-8 V) and high doping state ($V_G$=8 V). (c) Two-dimensional contour plot of photocurrent versus the gate voltage $V_G$ and the bias voltage $V_{DS}$. (d) The intrinsic responsivity R and internal quantum efficiency IQE as a function of applied bias for different thicknesses. $t_{BP}$: the thickness of BP. (a-d) are reproduced from Ref.[103] with permission from the Nature Publishing Group.

Recently, Bie et al. reported a silicon photonic-crystal (PhC) waveguide-integrated light source and photodetector based on a p–n junction of bilayer $MoTe_2$.[91] Figure 8a shows the device schematic, and Figure 8b shows LED mode and photodetector mode of device, respectively. The $MoTe_2$ p-n junction is composed of p-type and n-type doping that are electrostatically induced by left and right graphite top gates above the hBN dielectric layer. The bilayer $MoTe_2$ not only has stronger excitation and longer emission wavelength than the monolayer $MoTe_2$, but also has excitation caused by direct bandgap optical transitions similar to that of the monolayer.[74, 114, 115] There is a grating coupler for excitation and collection at the far end of the waveguide. Excitation light is coupled into the waveguide and transmitted along the waveguide to the grating coupler in LED mode, while incident light is coupled into the waveguide at the grating coupler and detected by the $MoTe_2$ p-n junction in the photodetector mode (Figure 8b). Figure 8c shows an EL emission intensity image at room temperature cover on top of a false-colour optical image of the device. Two extra emission spots are observed on the grating coupler on both sides, which coincides with the aforementioned LED mode (Figure 8b). The orange and green lines in Figure 8d represent the transmission spectra of the PhC waveguide before and after the bilayer $MoTe_2$ transfer, respectively. The orange curve in Figure 8d shows that the PhC transmission peak is at 1,160 nm, but this peak disappears after $MoTe_2$ transfer, which is mainly due to the interference of $MoTe_2$ absorption. The normalized EL emission spectrum collected at the grating coupler represented by the blue line in Figure 8d, showing a narrow peak centred at 1,160 nm on top of a broader peak centred at 1,175 nm.[91] The former coincides with the PhC transmission peak, and the latter matches the free-space emission of the bilayer $MoTe_2$. The reason that the PhC transmission peak can be observed at the blue line is that the excited states of $MoTe_2$ have been filled by the electrically excited carriers in the LED mode, so the absorption of $MoTe_2$ will not cause interference. When the device is operating in photodetector mode, the responsivity and external quantum efficiency of the photodetector at the wavelength of 1,160 nm are 4.8 mA/W and 0.5%, respectively. Figure 8e shows the temporal response in photodetector mode. The photocurrent response bandwidth is 200 MHz (limited by experimental set-up), which is much higher than other $MoTe_2$ photodetectors based on photogating and photoconductive effect,[116, 117] indicating the fast response of waveguide-integrated photodetector with bilayer $MoTe_2$ p-n junction. As can be seen from Figure 8f, the response spectrum of photodetector in the range from 1040 nm to 1240 nm and the largest responsivity is at 1,160 nm (the location of the PhC transmission peak), which shows the dependence of the responsivity on the wavelength. This dependence indicates that the photocurrent originates mainly from the light transmitted by the PhC waveguide, and thus the light beyond this spectral response range cannot be detected due to the bandgap of the bilayer $MoTe_2$ and the transmission cut-off of the waveguide. This work shows that all active optical components for point-to-point interconnects are possible with 2D TMD devices transferred onto otherwise passive photonic integrated circuits.[91]

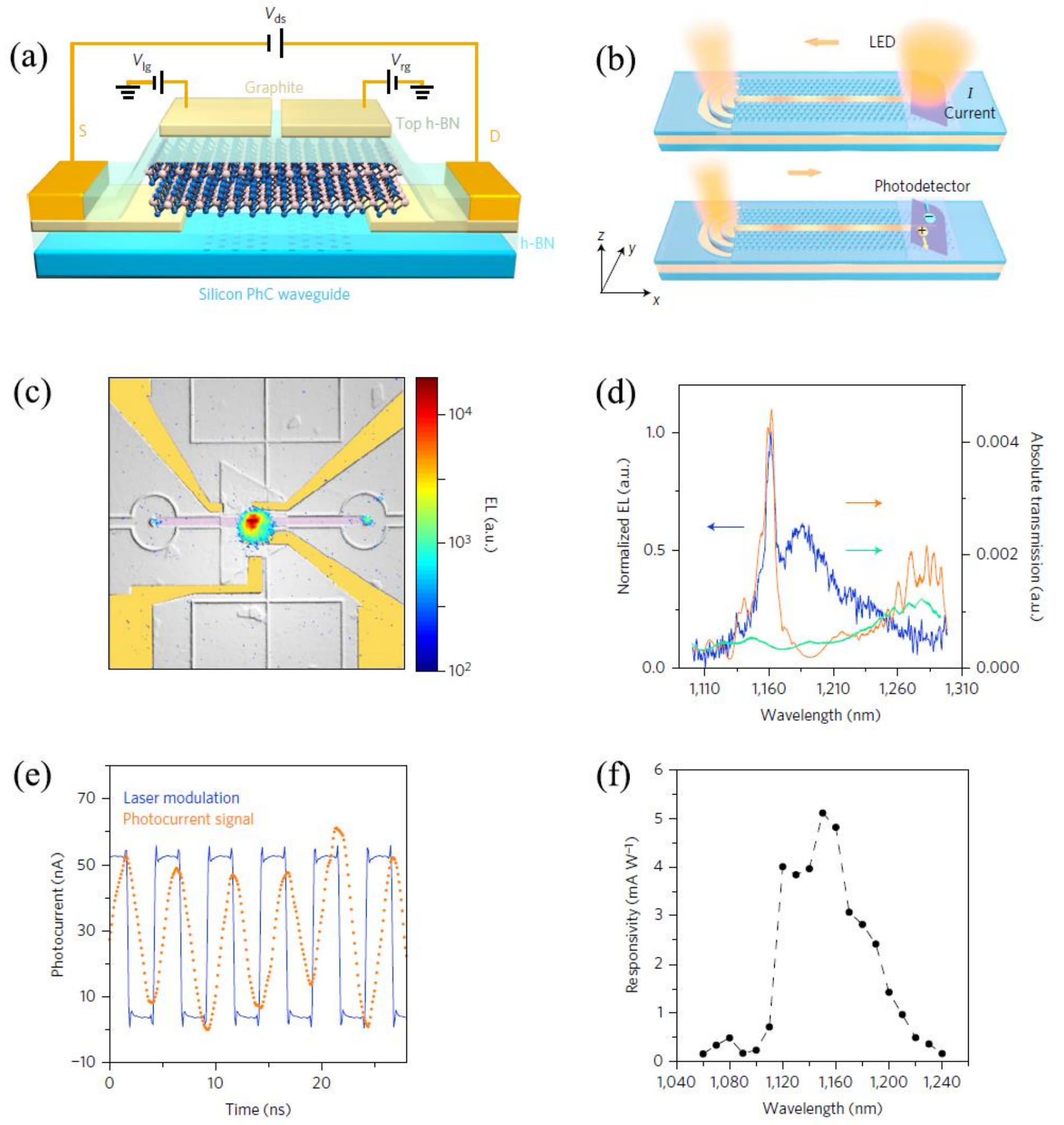

**Figure 8** (a) Schematic of the silicon photonic-crystal (PhC) waveguide-integrated light-emitting diode and photodetector based on bilayer $MoTe_2$ p–n junction. (b) The schematic diagram of the device operating in LED mode and photodetector mode. The arrow indicates the direction of light propagation in the PhC waveguide. (c) EL emission intensity image at room temperature cover on top of a false-colour optical image of the device. (d) The orange and green lines represent the transmission spectra of the waveguide before and after the bilayer $MoTe_2$ transfer, respectively. The normalized EL emission spectrum collected at the grating coupler represented by the blue line. (e) The temporal response of device in photodetector mode at zero bias. The response bandwidth of the device is 200 MHz. (f) The response spectrum of photodetector in the range from 1,040 nm to 1,240 nm, which shows the dependence of the responsivity on the wavelength due to photonic-crystal waveguide. (a-f) are reproduced from Ref.[91] with permission from the Nature Publishing Group.

### 3.2 2D materials on cavity

The integration of two-dimensional materials and cavities, including photonic crystal cavities,[118, 119] optical microcavities[120, 121] and optical ring resonators,[122] is also an effective method for enhancing the light absorption of 2D materials. The incident light at the resonant frequency of the cavity is absorbed many times by the atomically thin two-dimensional material in the cavity, thereby enhancing absorption. Furchi et al. achieved a 26-fold absorption enhancement of graphene (>60%) by monolithically integrating graphene with a Fabry-Pérot microcavity.[123] The schematic of the microcavity-integrated graphene photodetector and the corresponding electric field distribution are shown in Figure 9a. Two Bragg mirrors are alternately composed of quarter-wavelength thick materials with different refractive index, and graphene layer is sandwiched between these mirrors. The presence of the barrier layer ensures the position of resonance occur at the graphene. The incident light are reflected back and forth multiple times between two Bragg mirrors and absorbed by the graphene, getting a higher absorption, which agrees with the electric field distribution. The spectral response of the device is shown in Figure 9b, and it can be clearly seen that a significant photocurrent enhancement at the cavity resonance (855 nm). Atomic thin monolayer graphene (only 0.335 nm) can absorb more than 60% of light, indicating that the integration of 2D materials with cavity has great potential in enhancing the light absorption of 2D materials.

However, the enhancement of light absorption by the cavity occurs only at the resonant wavelength of the cavity. In other words, the enhancement of the cavity is at the expense of the broadband photodetection. The strategy of integrating cavity and two-dimensional materials cannot be used for broadband detectors, and thus it can only be used to enhance light absorption at specific wavelengths. For example, Casalino et al. demonstrated a vertically illuminated, resonant cavity enhanced, graphene−Si Schottky photodetector operating at communication band (1,550 nm).[124] Figure 9c shows the resonant cavity enhanced graphene/Si Schottky photodetector and the resonant structure consists of a λ/2 Si slab layer confined between SLG/Si top and Au bottom mirrors.[124] The response spectrum of the device with and without Au bottom mirror is shown in Figure 9d. A 3-fold external responsivity $R_{ext}$ enhancement as compared with no Au bottom mirror. The above result shows the integration of cavity with two-dimensional materials can achieve stronger and selective light absorption of 2D materials, so this method is possible to be applied to specific wavelength detection.

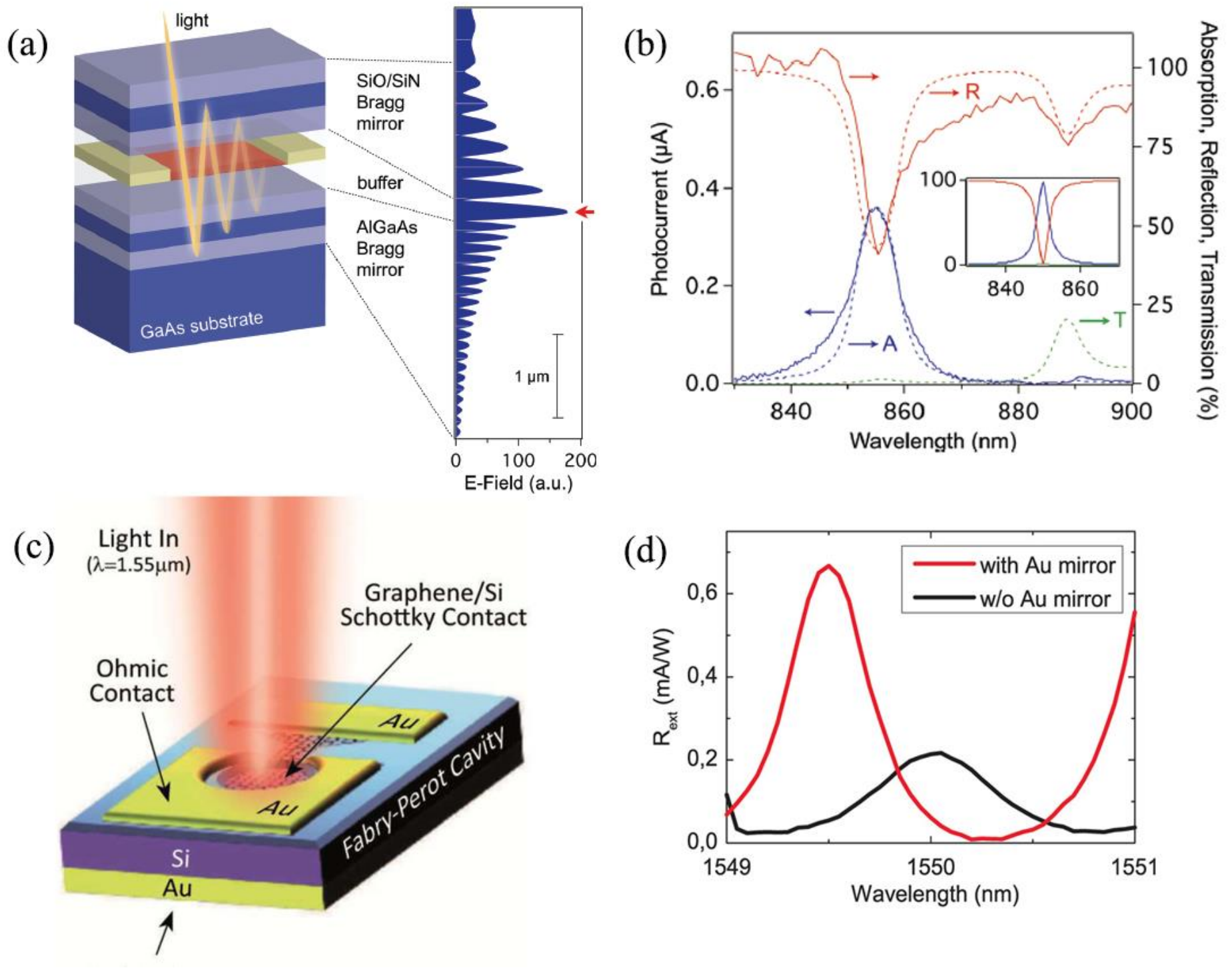

**Figure 9** (a) The schematic of the resonant cavity enhanced graphene photodetector and the corresponding electric field distribution. Red arrow: the position of graphene. (b) The dashed lines indicates the calculation result: reflection R (red), transmission T (green) and absorption A (blue), while the solid lines indicates the measurement result: reflection (red) and photocurrent (blue). A 26-fold absorption enhancement at the cavity resonance wavelength (855 nm). Inset: calculated theoretical results. (c) Schematic diagram of the resonant cavity enhanced graphene/Si Schottky photodetector. (d) The response spectrum of the device with and without Au mirror. A 3-fold absorption enhancement can be observed in the presence of Au mirror. The blue shift of the resonance peak is caused by the difference between different devices. (a-b) are reproduced from Ref.[123] with permission from the American Chemical Society. (c-d) are reproduced from Ref.[124] with permission from the American Chemical Society.

## 4. Plasmonic structure

In order to further enhance the photoresponse of the photodetector and obtain a tunable response spectrum, plasmonic nanostructures can also be applied to the infrared photodetector based on two-dimensional materials.[125-133] Metal plasmonic nanostructures mainly enhance the response of the detector in two ways. First, the near-field enhancement capability of the plasmonic nanostructures can focus the light field in a small area, effectively enhancing the light absorption of the two-dimensional material. Moreover, the plasmonic structure can generate plasma-induced hot electrons due to plasmon decay under light illumination, and then hot electrons are injected into the device across the Schottky barrier between the metal and the two-dimensional material to form photocurrent.

Fang et al. demonstrated a graphene-antenna sandwich photodetector with response spectrum in the range of visible to near-infrared and explained the reasons for photocurrent enhancement.[131] The schematic of photodetector and optical image are shown in Figure 10a and b. The Fano resonance plasma clusters (dimers, heptamers, and nonamers) sandwiched between two layers of graphene are respectively located in the 1, 3 and 5 regions, while the 2 and 4 regions have no plasmonic structure (Figure 10b). The Raman image in Figure 10b shows that the plasmonic antenna clusters of heptamers and nonamers significantly enhance Raman intensity due to resonance under 785 nm laser illumination. The photocurrent response in different regions in Figure 10c shows that the metal plasmonic antenna clusters can effectively enhance the photocurrent. As heptamers can provide larger near-field enhancement and higher thermal electron yield, the enhancement effect of heptamers is significantly better than that of the dimer. In addition, Figure 10d shows that the photocurrent enhancement of the dimer plasmonic clusters is related to the polarization of the incident light, indicating that this structure has potential in the field of infrared polarization detection. Figure 10e and f show the measured photocurrent and the corresponding band diagram for different applied gate voltages, respectively. The gate voltage $V_G$ can effectively adjust the Fermi level of the graphene channel, and then obtain different energy band bending between the Fermi level of graphene channel and the Ti electrode-doped graphene. The steeper band bending results in a correspondingly larger built-in electric field, which can obtain higher the separation efficiency of the photogenerated electron-hole pairs and thus produces a greater photocurrent. The resonance wavelength of the plasmonic clusters can be controlled by varying the diameter of disks. For example, when the diameter of heptamer plasmonic clusters changes from 80 nm to 180 nm, the corresponding resonance wavelength changes from 650 nm to 950 nm. Figure 10g shows the experimentally measured photocurrent as a function of different resonant wavelength (i.e. different heptamer diameters). It can be clearly seen that the photocurrent consists of plasmon-induced hot electrons and photogenerated carriers of plasmon-enhanced direct carrier excitation. The mechanism of photocurrent enhancement by plasmonic nanostructures is as follows. When the incident light resonates with the transmission window of the plasmonic clusters, a near-field enhancement effect occurs.[134] The light absorption of graphene is effectively enhanced and more electron-hole pairs are generated by direct excitation (DE). Moreover, hot electrons (HE) generated in metal nanostructures can transfer into the conduction band of graphene over the Schottky barrier between metal and graphene.[135] Both plasmon-induced hot electrons and directly excited carriers can be driven by the source-drain bias into the circuit to from photocurrent. Because the contribution of absorption relative to scattering decreases as the size of the nanoparticles increases, the contribution of HE to photocurrent tends to saturate, while the contribution of DE plays a greater role with increasing the diameter of the heptamer.[131]

Since hot electrons are generated by plasmon decay, the response spectra of photodetectors based on two-dimensional materials and metal plasmonic nanostructures are not limited by the bandgap of two-dimensional materials.[129, 130] For example, bilayer $MoS_2$ has an indirect bandgap of 1.65 eV, corresponding to a wavelength of 750 nm.[136] The hot electron-based bilayer $MoS_2$ photodetector has a high responsivity of 5.2 A/W at 1,070 nm (Figure 10h).[129] $MoS_2$ field-effect phototransistors were also found to have negative infrared photoresponse, which is caused by bolometric effect, and exhibited responsivity of 2.3 A/W at 980 nm.[137] The above results show that plasma-induced hot electrons can achieve effective light detection with below bandgap photon illumination.

Ni et al. reported a photodetector based on plasmonic Si QD/graphene heterostructure that enables effective detection from UV to mid-infrared.[138] B-doped Si QDs not only enhance the light absorption of graphene in the infrared region due to surface plasmon resonance but also form photogating effect, which greatly improves the performance of the detector. Thus, the device exhibits ultrahigh gain of ~$10^{12}$ and responsivity of ~$10^9$ A/W.

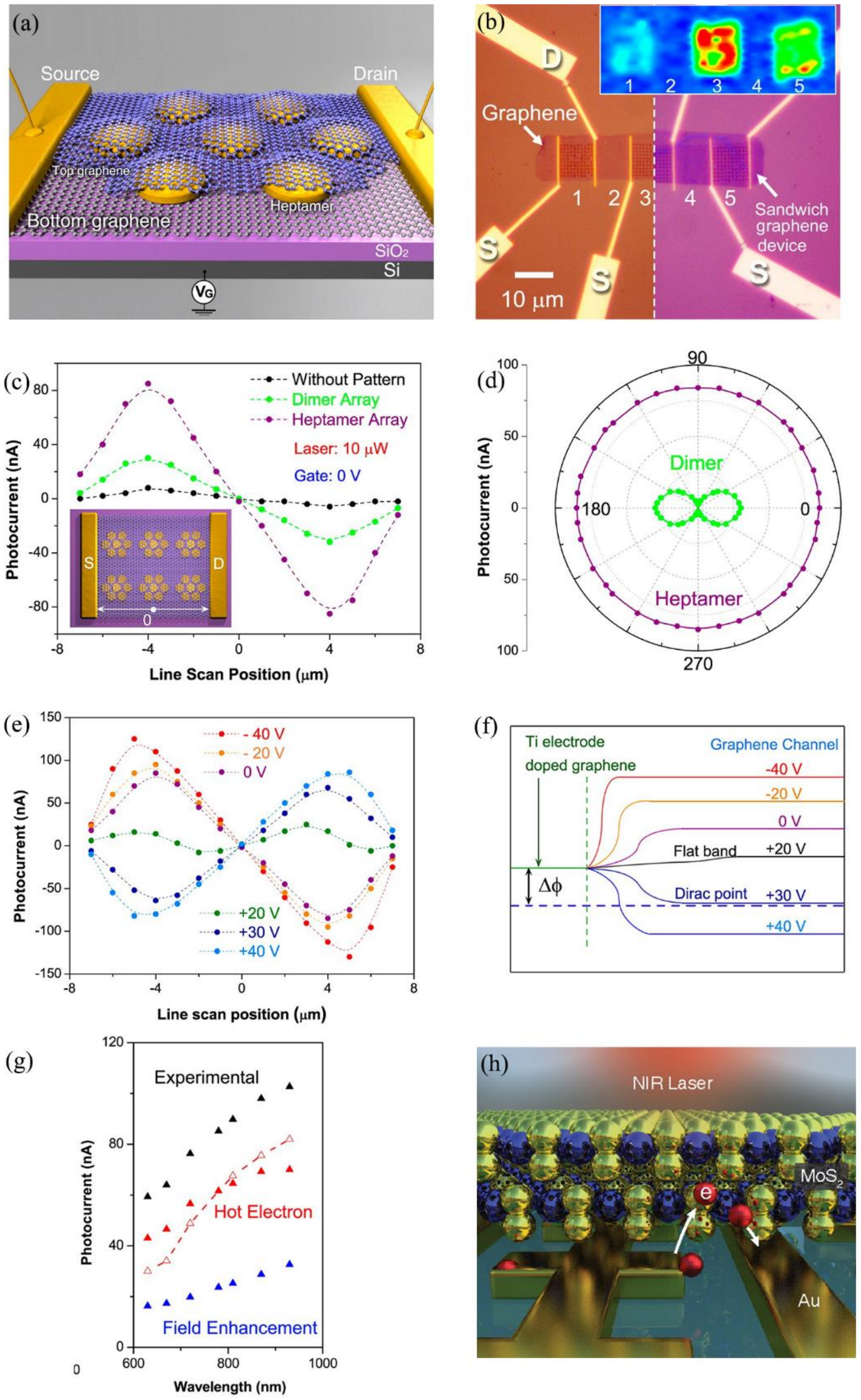

**Figure 10** (a) The schematic of graphene-antenna sandwich photodetector. (b)

The optical image of device before (left) and after (right) transfer of top layer graphene. Inset: Raman mapping of the device under 785 nm laser illumination. (c) Photocurrents measured in the photodetector with dimer array and heptamer array and without pattern exhibit antisymmetric properties. The results show that the metal plasmonic antenna clusters can effectively enhance the photocurrent. The line scan position is shown by the white arrows in the inset. (d) The polarization dependent properties of the photocurrent measured in dimer and heptamer devices. (e) Photocurrents measured at different gate voltages in a heptamer device. (f) Energy band diagram of devices with different gate voltages in a heptamer device. The purple dashed line indicates the Dirac point of graphene. (g) Photocurrent as a function of different resonant wavelength. Black triangles: the measured photocurrent ($I^{EXP}(\lambda)$). Blue triangles: the calculated DE photocurrent ($I^{DE}(\lambda)$). Solid red triangles: the estimated HE photocurrent contribution $E^{DE}(\lambda) = I^{EXP}(\lambda) - I^{DE}(\lambda)$. Hollow red triangles: the calculated $I^{HE}(\lambda)$. (h) Schematic of hot electron-based bilayer $MoS_2$ with near-infrared laser illumination. (a-g) are reproduced from Ref.[131] with permission from the American Chemical Society. (h) is reproduced from Ref.[129] with permission from the American Chemical Society.

Table 1. Performance of room temperature infrared photodetectors with hybrid structure based on 2D materials. (UV: Ultraviolet. NIR: Near infrared. MIR: Mid-infrared. BP: Black phosphorus. Gr: Graphene. QD: Quantum dot. FET: Field effect transistor.)

| Device type | Device | Spect | Responsi | Detectivi | Respon | EQE/I | gain | Ref. |
|---|---|---|---|---|---|---|---|---|

| | structure | ral range | vity (A/W) | ty (Jones) | se time (s)/$f_{3dB}$ bandwidth | QE | | |
|---|---|---|---|---|---|---|---|---|
| Hybrid 2D-0D structures | Gr-PbS QD | NIR | $10^7$ | - | 0.26 | - | - | [16][16] |
| | Gr-PbS QD hybrid | NIR | $1.6\times10^4$ | $8.6\times10^{10}$ | $8\times10^{-3}$ | $2.4\times10^6$% | - | [30] |
| | $MoS_2$-HgTe QD | Visible-NIR | $10^6$ | $10^{12}$ | $\sim10^{-3}$ | - | $10^6$ | [31] |
| | Gr/PbS QD | Visible-NIR | $\sim10^7$ | $7\times10^{13}$ | 0.01 | 25% | $\sim10^8$ | [32] |
| | Gr/PbS QDs/Gr | Visible-NIR | 58 | $2\times10^{11}$ | ~0.03 | - | - | [33] |
| | ITO/PbS QD/Gr | Visible-NIR | $2\times10^6$ | $10^{13}$ | 1.5 KHz | 70%-80% | $10^5$ | [34] |
| | $MoS_2$/PbS QD | Visible-NIR | $10^5$-$10^6$ | $5\times10^{11}$ | 0.3-0.4 | - | - | [35] |
| | $MoS_2$/PbSe QD | NIR | 1.9×10-6 | - | 0.25 | - | - | [36] |
| Hybrid 2D-1D structures | SWNT-Gr hybrid films | 400-1,550 nm | >100 (650 nm) ~40 (1550 nm) | - | $1\times10^{-4}$ (650 nm) | 34% (650 nm) | $10^5$ (650 nm) | [50] |
| | SWNT-$MoS_2$ | Visible-NIR | >0.1 (650 nm) | - | $<1.5\times10^{-5}$ (650 nm) | 25% (650 nm) | - | [60] |
| | $WSe_2$-$In_2O_3$ | Visible-NIR | $7.5\times10^{5}$ (637 nm) $3.5\times10^{4}$ (940 nm) | $4.17\times10^{17}$ (637 nm) $1.95\times10^{16}$ (940 nm) | $2\times10^{-2}$ (940 nm) | - | - | [139] |
| | Gr-$WSe_2$-Gr | NIR | $0.12\times10^{-3}$ (1500 nm) | - | - | 2% | - | [140] |
| | $MoS_2$ PN homojunction | Visible-NIR | $7\times10^{4}$ | $3.5\times10^{14}$ | $\sim10^{-2}$ | >10% | $>10^{5}$ | [141] |
| | Gr-$Bi_2Se_3$ | Visible-MIR | 1.97 (3500 nm) 8.18 (1300 nm) | $1.7\times10^{9}$ | $4\times10^{-6}$ | 2.3% | 30 | [142] |
| Hybrid 2D-2D structures | Few-Layer α-$MoTe_2$/$MoS_2$ | Visible-NIR | 0.322 (470 nm) 0.037 (800 nm) | - | $2.5\times10^{-2}$ (620 nm) | 85% (470 nm) 6% (800 nm) | - | [76] |
| | $MoS_2$-Gr-$WSe_2$ | Visible-MIR | 0.306 (940 nm) | $10^{11}$ (940 nm) | $3.03\times10^{-5}$ (637 nm) | $10^{6}$% (532 nm) | - | [79] |
| | Gr-$Bi_2Te_3$ | UV-NIR | ~10 (980 nm) | - | $9.3\times10^{-3}$ (1550nm) | - | 83 (532 nm) | [82] |
| | BP/$WSe_2$ | Visible-NIR | $10^{3}$ (637 nm) 0.5 | $10^{14}$ (637 nm) $10^{10}$ | $8\times10^{-4}$ | - | $10^{6}$ (637 nm) | [85] |
| | | | (1550 nm) | (1550 nm) | | | $10^2$ (1550 nm) | |
| | BP p-n junction | NIR | 0.18 | ~$10^{11}$-$10^{13}$ | 0.015 | 0.75% | - | [88] |
| Waveguide-integrated structures | $MoTe_2$ p-n junction | NIR (1110 -1200 nm) | $4.8\times10^{-3}$ (1160 nm) | - | 200 MHz | 0.5% (1160 nm) | - | [91] |
| | metal-doped Gr junction | NIR | 0.1 | - | 20 GHz | 3.8% | - | [98] |
| | Metal-Gr-Si | 1550 nm | 0.37 | - | - | 7% | 2 | [99] |
| | Gr-$MoTe_2$-Au | NIR | 0.023 (19.5 nm) 0.4 (60 nm) | - | ~1 GHz | 1.5% (19.5 nm) 35% (60 nm) | - | [100] |
| | Gr | NIR | 0.05 | - | ~18 GHz | 10% | - | [101] |
| | Gr/Si | Visible-MIR | 0.13 (2750 nm) | - | $4\times10^{-8}$ | 71.5% | - | [102] |
| | BP FET | NIR | 0.135 (11.5-nm-thick) 0.657 (100-nm-thick) | - | >3 GHz | 10% (11.5-nm-thick) 50% (100-nm- | - | [103] |
| | | | | | | thick) | | |
| Microcavity-integrated structures | Gr | 864.5 nm | $2.1\times10^{-2}$ | - | - | - | - | [123] |
| | Gr-Si | 1550 nm | $2\times10^{-2}$ | $5.1\times10^{7}$ | $1.35\times10^{-9}$ | - | - | [124] |
| Plasmonic-integrated structures | Pt nanostrips - $MoS_2$ | UV-NIR | 14 (325 nm) 312.5 (532 nm) 69.2 (980 nm) | $1.113\times10^{10}$ (532 nm) | 0.707 (532 nm) | $7.283\times10^{4}$ (532 nm) | - | [126] |
| | Gr-Au nanorods | UV-NIR | $4\times10^{4}$ (1310 nm) | - | 0.7 s (1310 nm) | - | $1.8\times10^{9}$ (1310 nm) | [128] |
| | Bilayer $MoS_2$-Au | Visible-NIR | $1.1\times10^{5}$ (532 nm) 5.2 (1070 nm) | - | 28.5 (532 nm) 44.5 (1070 nm) | - | $1.05\times10^{5}$ (532 nm) | [129] |
| | Few-layer $MoS_2$-metal junctions | Visible-NIR | $\sim4\times10^{-3}$ (650 nm) $\sim0.5\times10^{-4}$ (850 nm) | - | - | - | - | [130] |
| | Gr-antenna | Visible-NIR | $\sim1.25\times10^{-2}$ | - | - | 22% | - | [131] |

## 5. Conclusion and prospects

In this review, we have reviewed various infrared photodetectors operating at room temperature with hybrid structure based on two-dimensional materials. Hybrid 2D-QDs heterostructures possess two prominent advantages of strong light absorption contributed by quantum dots and high carrier mobility contributed by two-dimensional materials, and thus exhibiting extremely high gain and responsivity. The detection performance of photodetectors based on hybrid graphene-SWNTs structure is superior to that of photodetectors based on only graphene or SWNTs, showing high responsivity and fast response, in which the bandgap of SWNTs can be adjusted by the diameter of SWNTs to extend the spectral response range to infrared. The 2D-2D van der Waals heterostructure is also suitable for light detection in the infrared region because of ultrafast charge transfer at the interface, and the combination of two or more two-dimensional materials with different optical and electrical properties not only compensates for the lack of both (eg. the combination of 2D materials with different bandgap structures can achieve broadband detection), but also exploits the characteristics of 2D materials to prepare a photodetector with specific function (such as high polarization sensitivity of a BP-on-$WSe_2$ photogate structure). The integration of two-dimensional material on waveguide makes the light-matter interaction no longer limited by the ultrathin thickness of two-dimensional material, greatly increasing the light absorption of photodetectors to obtain a higher responsivity. In addition, this integration also shows that two-dimensional materials are compatible with the existing mature optical waveguide technology, providing a pathway for the commercialized practical application of two-dimensional materials. The integration of two-dimensional material with the cavity enhances the light absorption of the material at the resonant wavelength of the cavity. This selective absorption method is very suitable for precise detection of specific wavelengths. Metal plasmonic nanostructures can also enhance light absorption and photocurrent through near-field enhancement effects and hot electrons. In addition, plasma-induced hot electrons can extend the spectral response range, which is useful for infrared detection based on some two-dimensional materials with wide bandgap.

Infrared photodetectors with hybrid structure based on two-dimensional materials still face many challenges and need us to explore further.

1) **Low-cost, large-scale preparation**

Many of the currently reported hybrid structures based on two-dimensional materials are prepared just only through mechanical exfoliation and targeted transfer techniques, leading to poor reproducibility and relatively low yield of photodetectors. This method can only meet the needs of scientific research and cannot be used for commercial applications. Although the 2D vertical or lateral heterostructure can be prepared by the CVD method and 2D-QDs structure can be prepared by the solution-processed method, which shows the potential of large-scale production, but they still have many limitations and need further improve.

2) **Device performance optimization**

The reported overall performance parameters of infrared photodetectors with hybrid structure based on two-dimensional materials are excellent, but they are still far from the theoretical value. The excellent optical and electrical properties of two-dimensional materials have not been fully exploited. In order to obtain better detection performance, it is necessary to further study the interfacial charge transfer and photocurrent generation mechanism in hybrid structure based on two-dimensional materials for designing or optimizing various hybrid structures.

3) **Research on novel and suitable two-dimensional materials**

There are various currently limitations for two-dimensional materials used in the field of infrared detection. For example, the band gap of graphene and many TMDs (such as $MoS_2$, $WSe_2$) are not conducive to the absorption of infrared light. The instability of $MoTe_2$ and BP with an appropriately bandgap in the air also limits their use. In order to obtain better infrared detection performance, novel two-dimensional materials with suitable band gap and good stability have to be developed.

4) **Design of novel heterostructure with 2D materials**

Many existing heterostructures based on two-dimensional materials simply combine two materials that appear to be suitable. However, this combination may not be optimal, which may leads to the inability to maximize the advantages of heterostructure. It is a worthwhile research direction to further understand the properties of various two-dimensional materials through theoretical research and computational simulation methods, and to design a more rational and ingenious novel heterostructure. It has the potential to dramatically improve the performance of the photodetector and even discover new phenomena.

5) **Interface research of heterostructures**

Two-dimensional materials are particularly sensitive to interface problem due to their large specific surface area and atomic thickness. On the one hand, interface pollution and the introduction of defects have a huge impact on the physical properties of two-dimensional materials. Therefore, how to obtain a clean and defect-free interface in the preparation process of heterostructure based on two-dimensional materials has always been a research hotspot. On the other hand, the study of charge transfer and interaction at the interface of two materials also helps to clarify the working mechanism of the photodetector and to improve performance of photodetector after the formation of the heterostructure.

**Declaration of conflicts of Interest**

The authors declare no conflict of interest.

**Acknowledgements**

The authors acknowledge funding by National Natural Science Foundation of China (No. 61704061)